# Review and Analysis of Measurements of the Spin Hall Effect in Platinum

Luqiao Liu[1], R. A. Buhrman[1], and D. C. Ralph[1,2]
[1]Cornell University and [2]Kavli Institute at Cornell, Ithaca, New York, 14853

## Abstract

Several different experimental techniques have been used in efforts to measure the spin Hall conductivity $\sigma_{SH}$ and the spin Hall angle $\theta_{SH}$ in Pt samples at room temperature, with results that disagree by more than a factor of 20, from $\sigma_{SH} = 2.4 \times 10^4 [\hbar/(2e)](\Omega m)^{-1}$ to $5.1 \times 10^5 [\hbar/(2e)](\Omega m)^{-1}$ and from $\theta_{SH} = 0.0037$ to $0.08$. We review this work, and analyze possible reasons for the discrepancies. We explain that the smallest values for $\sigma_{SH}$ that have been reported, based on measurements of lateral permalloy/copper/platinum devices, are incorrect because the original analyses did not properly take into account that copper layers in these devices will shunt charge current flowing through adjacent platinum wires, thereby greatly reducing the size of the spin-Hall-related signals. We suggest that differences between the results for $\sigma_{SH}$ found by other experimental techniques are primarily a consequence of different assumptions about the value of the spin diffusion length, $\lambda_{sf}$. We present a new measurement of $\lambda_{sf}$ for Pt within sputtered Pt/permalloy bilayer thin films at room temperature, finding $\lambda_{sf} = 1.4 \pm 0.3$ nm, a much smaller value than has generally been assumed previously. With this value for the spin diffusion length, the previously-discordant results can be brought into much better agreement, with the result that $\sigma_{SH} = (1.4 - 3.4) \times 10^5 [\hbar/(2e)](\Omega m)^{-1}$ and $\theta_{SH} > 0.05$. These values are sufficiently large that the spin Hall effect in Pt can be used to generate spin currents and spin transfer torques strong enough for efficient manipulation of magnetic moments in adjacent ferromagnetic layers.

**Corrections in this version of the manuscript, relative to the version posted originally.**

1. Our original criticism was incorrect regarding a procedure introduced in 2011 by the Otani group to correct for current shunting in lateral permalloy/copper/platinum devices. After correspondence with Prof. Otani and our own independent calculations, we now agree that this correction procedure should be a good approximation for samples in which the spin diffusion length is longer than the film thickness of the spin Hall metal. However, all of the spin Hall metals (including Pt) studied by the Otani group in ref. [14] are reported to have spin diffusion lengths less than the film thickness, and for this case we continue to believe that the correction procedure used by Prof. Otani and collaborators can lead to significant underestimates of the spin Hall angle, underestimates that are the most severe for spin Hall metals with the largest resistivities. See the revised discussion on pages 6-12. We are grateful to Professors Otani and Fert for their correspondence.

2. We note on page 6 that H. Nakayama *et al.* (IEEE Trans. Magn. **46**, 2202 (2010)) explained the importance of correcting for the shunting effect when interpreting inverse spin Hall effect measurements.

3. We have clarified on page 15 and endnote [23] our discussion of the Elliot-Yafet mechanism for spin relaxation. We mention this mechanism only to make the point that one should not assume that the spin diffusion length is the same in all Pt samples, because the spin diffusion length may depend on parameters such as the resistivity. We believe that it is still an open question how well the simple Elliot-Yafet predictions describe the spin diffusion length in Pt.

4. Two notes are added on page 28.

*******************

The purpose of this paper is to attempt to understand and reconcile previous measurements of the spin Hall conductivity $\sigma_{SH}$ and spin Hall angle $\theta_{SH}$ in Pt at room temperature, which have reported results that disagree by more than a factor of 20. Our hope is to generate discussion that might eventually lead to a consensus regarding the correct interpretation of the experiments. We therefore invite criticisms of our analysis, and we will revise this arXiv posting as necessary to correct any errors or to include new insights (with appropriate attribution).

The spin Hall angle $\theta_{SH}$ describes the magnitude of both the direct spin Hall effect and the inverse spin Hall effect [1-3]. In the direct spin Hall effect (DSHE), an applied charge current density $J_e$ flowing through a heavy metal such as Pt generates a transverse-flowing spin current density $(\hbar/2)J_S/e$ with a magnitude described by the spin Hall angle as $J_S = \theta_{SH}J_e$. In the inverse spin Hall effect (ISHE), an applied spin current density generates a transverse-flowing charge current density whose magnitude can be described by the same spin Hall angle, $J_e = \theta_{SH}J_S$ [3,4]. It is also useful to define the spin Hall conductivity, $\sigma_{SH}$, which is related to $\theta_{SH}$ according to $\sigma_{SH} = [\hbar/(2e)]\theta_{SH}\sigma$, where $\sigma$ is the charge conductivity of the material. The direct and inverse spin Hall effects have been the subject of intense study in semiconductors (*e.g.*, [5]) and metals (*e.g.*, [4,6,7,8]). The spin Hall effect can arise from either intrinsic (band structure) or extrinsic (scattering) effects in materials within which there is significant spin orbit coupling (for reviews see [9,10]). In general, in the intrinsic limit it is expected that the spin Hall conductivity is largely independent of the mean free path for scattering (and hence also of the electrical conductivity of the material), so that in the intrinsic limit for a given type of material the spin Hall angle should scale as $\theta_{SH} \propto 1/\sigma$. In an extrinsic limit where skew scattering processes are dominant, one should have $\sigma_{SH} \propto \sigma$ and the spin Hall angle $\theta_{SH}$ should be approximately independent of scattering within the material.

In this manuscript we will focus primarily on experiments on Pt, because this is the most commonly measured spin-Hall material and it is the material for which at present there is the largest range of conflicting results. We will keep in mind that the value of $\theta_{SH}$ need not be the same for all Pt samples, *e.g.*, if the intrinsic limit applies then one should expect $\theta_{SH} \propto 1/\sigma_{Pt}$. For this reason we will be careful to note the conductivity of the Pt films for each experiment that we discuss. However, the differences between the results reported for different experiments are far too large (and often in the wrong direction) to be explained simply by this simple dependence of $\theta_{SH}$ on $\sigma_{Pt}$.

To our knowledge the first attempt to make a quantitative measurement of $\theta_{SH}$ for Pt was reported in 2007 by Kimura *et al.* from the Otani group [7]. This paper described measurements of both the direct and inverse spin Hall effects in the same device, in which the electrical conductivity of the Pt was $\sigma_{Pt} = 6.4 \times 10^6$ $(\Omega m)^{-1}$. In the first case, the authors used the direct spin Hall effect to inject a spin current from a Pt wire into a Cu wire, and measured the resulting spin accumulation in the Cu using a permalloy contact. In the second case, they injected a nonlocal spin current from the permalloy through the copper to the Pt wire, and measured the transverse voltage generated across the Pt due to the inverse spin Hall effect. In both sets of experiments, the original analysis of the signals indicated a spin Hall angle of 0.0037, corresponding to $\sigma_{SH} = 2.4 \times 10^4 [\hbar/(2e)](\Omega m)^{-1}$. This work was pioneering in that it was the first experiment to demonstrate the direct spin Hall effect in a metal, and it also showed that the spin Hall conductivity in Pt is much larger than in semiconductors and aluminum (the materials measured previously). However, the results are not quantitatively correct because the original analysis of the data did not take into account that the Cu wire in contact with the Pt will shunt the Pt, altering the results of both the direct and inverse spin Hall experiments. The difficulty is illustrated in Fig. 1(a), in which the arrows indicate the direction and relative magnitude of the local charge density near a Cu/Pt interface when a voltage is applied to the ends of the Pt wire. The original analysis of the direct spin Hall measurement assumed that the applied charge current flowed only within the Pt wire. However, since Cu has a larger electrical conductivity than Pt (ref. [7] gives $\sigma_{Cu} = 48 \times 10^6$ $(\Omega m)^{-1}$ compared to $\sigma_{Pt} = 6.4 \times 10^6$ $(\Omega m)^{-1}$) and the Cu wire is also much thicker than the Pt, a majority of the charge current will be shunted through the Cu instead, and the charge current density in the Pt near the Cu interface will be greatly reduced. This decrease in the charge current flowing in the Pt will reduce the spin accumulation in the Cu arising from the spin Hall effect by a factor $x_S$ (relative to the value under the assumption that all the charge current flows within the Pt) and will therefore cause an underestimate of $\theta_{SH}$ by the same factor.

The Cu overlayer will affect the inverse spin Hall effect measurement similarly. In this case, a spin current injected from the Cu into the Pt gives rise to a transverse charge current flowing in the Pt. The experiment of Kimura *et al.* [7] measured the voltage generated across the Pt wire under an open circuit condition that no net charge current flowed to the external circuit. The measured voltage in this case is therefore determined by the condition that a backward-flowing current associated with the voltage cancels the forward-flowing current generated by the inverse spin Hall effect. (We will describe how to perform this calculation below.) In the original analysis performed by Kimura *et al.*, it was assumed that the backward-flowing current arising from the voltage would flow only within Pt wire. However, for the device geometry studied, the voltage will actually drive backflow charge current in both the Pt and the Cu overlayer, which means that much less voltage is required to cancel the forward-going inverse spin Hall effect current than would be the case if no charge current flowed within the Cu. (The need to correct for this shunting effect in inverse spin Hall experiments was pointed out by references [11] and [12]). A quantitative analysis shows that the shunting decreases the inverse spin Hall signal by the same factor $x_S$ as for the direct spin Hall effect experiment, leading to the same underestimate of $\theta_{SH}$ in the inverse spin Hall effect measurement as for the direct spin Hall effect measurement. This must be the case in order for the Onsager reciprocity relationships discussed by Kimura *et al.* to be obeyed. We believe that the neglect of this shunting effect is therefore sufficient to explain why the values of $\sigma_{SH}$ ($2.4 \times 10^4 [\hbar/(2e)](\Omega m)^{-1}$) and $\theta_{SH}$ (0.0037) claimed by Kimura *et al.* are more than an order of magnitude smaller than measurements performed later by other techniques.

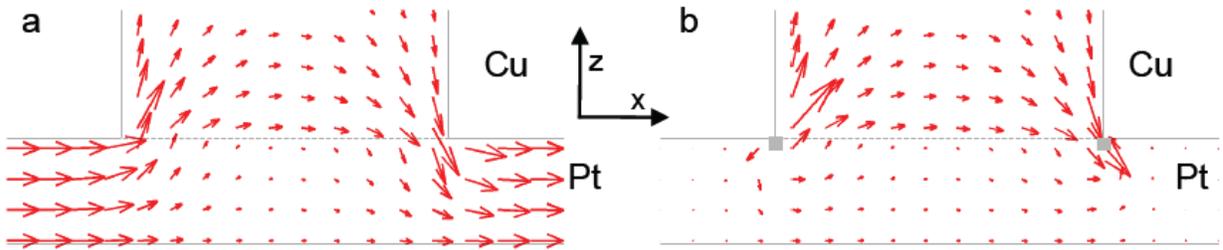

*Fig. 1*. *(a) Calculated charge current distribution in a Cu/Pt device showing current shunting by a Cu overlayer above a Pt wire when current flow is generated by application of a voltage between the ends of the Pt wire ($\nabla \cdot J_\varphi = 0$ inside the sample in this case). Only a part of the Cu overlayer is shown. (b) Calculated backflow charge current distribution generated by the inverse spin Hall effect when the spin diffusion length is 2 nm ($\nabla \cdot J_\varphi = -\nabla \cdot J_{ISHE}$ in this case). In both panels, each arrow represents the value of the charge current density at the position of the arrow's base, not its midpoint. We assume that the Pt layer is 20 nm thick and the Cu overlayer is 100 nm long and 100 nm thick. The shaded regions in (b) show approximately where the source term $\nabla \cdot \vec{J}_{ISHE}$ is nonzero when $\lambda_{sf}$ is 2 nm. We use $\sigma_{Pt} / \sigma_{Cu} = 0.13$. (For definitions of terms, see the discussion in the text.)*

The Otani group and collaborators have acknowledged the importance of this shunting effect, and in 2011 began to apply a correction within their analyses in an effort to account for the shunting (see the supplemental material for ref. [13] and ref. [14]). Their updated values for the spin Hall conductivity and spin Hall angle in Pt (measured at 10 K rather than at room temperature on samples with a 10 K conductivity $\sigma_{Pt} = 8.1 \times 10^6$ $(\Omega m)^{-1}$ ) are $\sigma_{SH} = 1.7 \times 10^5 [\hbar/(2e)](\Omega m)^{-1}$ and $\theta_{SH} \approx 0.021$ [14]. However, these values are still somewhat less than the values we will argue to be accurate based on measurements by other techniques, discussed below, and furthermore the spin Hall conductivity reported for the higher-resistivity spin Hall metal Ta in ref. [14] is in even more severe disagreement with recent measurements by our group [15]. (We measure the spin Hall angle in Ta to be even larger than in Pt, while ref. [14] reports a value 6 times smaller. We therefore expect that the analyses of the Ta experiments will be a contentious subject, as well.) We believe that the reason for the remaining discrepancies is that the correction procedure introduced in references [13] and [14] is not fully correct in how to account for the effects of shunting. We explain the reasons for our disagreement in the following paragraphs.

The correction procedure introduced in references [13,14] by the Otani group determines the fraction of the lateral component of the charge current that flows in the spin Hall metal (rather than the Cu shunt) compared to the total current through the bilayer device when a voltage is applied between the two ends of the Pt wire. (To determine the correction factor, the lateral component of the charge current in the spin Hall metal is averaged over the lateral direction and summed over the full film thickness in the vertical direction.) The Otani group calls this correction factor $x$; we will call it $x_e$. References [13] and [14] argue that the shunting-induced reduction in the size of the spin Hall and inverse spin Hall signals, $x_S$, should be quantitatively the same as the fraction of the total charge flow remaining in the Pt, or $x_e = x_S$. We agree that this should be a reasonable approximation in the limit that the spin diffusion length is much greater than the film thickness of the spin Hall metal, so that the entire thickness of the spin Hall film contributes equally to the direct spin Hall or inverse spin Hall signal (depending on which is measured). However, the thickness of the spin Hall metal layers in ref. [14], $t_{SHM}$ = 20 nm, is greater than the stated values of the spin diffusion lengths $\lambda_{sf}$ for all of the materials studied in that paper. Furthermore, our measurements indicate (see the discussion later in this review and also [16]) that $\lambda_{sf}$ in Pt and Ta is on the order of 1 nm, even shorter than the values stated in ref. [14]. For materials with $\lambda_{sf} \leq t_{SHM}$, the parts of the sample at different depths will not all contribute to the spin Hall experiments in the same way, and this complicates the shunting analysis. By the Onsager reciprocity principle, the correction factors should be the same for both the direct spin Hall and inverse spin Hall experimental configurations, but we find it instructive to analyze these two cases separately.

In the case of a direct spin Hall effect experiment, an external voltage is applied between the ends of the spin-Hall-metal wire so that the current distribution near the Cu interface should have a form similar to that in Fig. 1(a). Note that this current distribution varies with position both in the lateral and vertical directions. In the vertical direction, because of shunting by the Cu the current density in the spin Hall metal near the Cu interface can be significantly smaller than the charge current density averaged over the full thickness of the spin-Hall-metal film. When computing the strength of the spin current that will be injected into the Cu by the spin Hall effect, only the lateral charge current density flowing in the spin Hall metal within a distance from the Cu of approximately the spin diffusion length will contribute; the charge current density flowing near the bottom of the spin Hall metal wire will be immaterial. Consequently, when $\lambda_{sf}$ is short the reduction in the strength of the spin Hall signal is more severe than indicated by the shunting-induced correction factor $x_e$ that describes the reduction in the total current averaged over the full thickness of the spin-Hall-metal wire, i.e. $x_S < x_e$ when $\lambda_{sf} < t_{SHM}$. In Fig. 2 we plot calculations of the lateral component of the charge current density in the spin Hall metal, averaged over lateral positions within the shunting region, as a function of vertical position in the spin Hall metal, for the cases corresponding to the resistivities of Pt and Ta [15] (both with a Cu shunt). [We use $\sigma_{Pt}/\sigma_{Cu} = 0.13$ and $\sigma_{Ta}/\sigma_{Cu} = 0.005$, corresponding to the values in ref. [14].] These results suggest that the local charge current density near the Cu interface in the case of Pt/Cu is about 50% of the current density averaged over the full thickness of the Pt wire and in the case of Ta/Cu it is approximately only 10% of the full-thickness averaged charge density (the difference between materials being due to the much lower conductivity of Ta). If (as we will argue) the spin diffusion length is on the scale of 1 nm for both Pt and Ta, it is the charge current density very close to the Cu interface that contributes to the direct spin Hall effect, not the value averaged over the full film thickness.

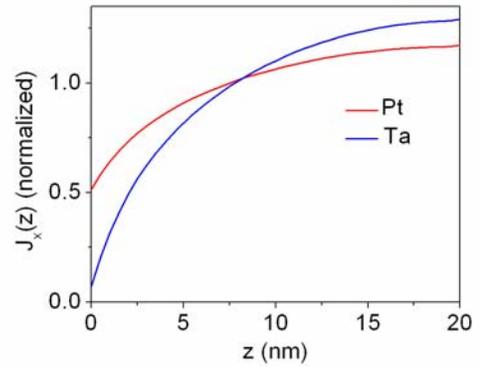

*Fig. 2. For a direct spin Hall effect experiment, the lateral component of the current density $J_x$ in the spin Hall metal (Pt or Ta) averaged over the x direction and plotted as a function of the vertical distance z from the interface with Cu (z = 0 is the interface position). The values are normalized relative to the average value $\langle J_x(z) \rangle$ over the whole thickness of the spin Hall metal film for each material.*

In the case of an inverse spin Hall effect experiment, the correction procedure used in references [13,14] is (i) to calculate the total charge current generated in the spin-Hall-metal wire by the inverse spin Hall effect, taking into account that this may be nonuniform in the vertical direction because of the finite spin diffusion length and (ii) calculate how much voltage must be applied between the ends of the spin-Hall-metal wire to produce a backflow current in the total bilayer that cancels (globally) the current generated by the inverse spin Hall effect, so that the overall net current leaving the device is zero. That is, this procedure assumes that the backflow current distribution may be calculated *independently* of the spatial distribution of the current density arising from the inverse spin Hall effect. This assumption is, however, incorrect because the two portions of the charge current density – the part arising from the spin Hall effect ($J_{ISHE}$) and the part associated with variations in the electric potential ($J_\varphi = -\sigma \nabla \varphi$) – are not independent; they are coupled through the steady-state equation of continuity for the total charge current density, $\nabla \cdot (J_{ISHE} + J_\varphi) = 0$, that must hold locally everywhere within the sample. The charge current distributions determined in references [13] and [14] do not satisfy the continuity equation because the solutions for $J_\varphi$ are based on the assumption that $\nabla \cdot J_\varphi = 0$ everywhere within the sample volume, despite the fact that $\nabla \cdot J_{ISHE}$ is explicitly nonzero in the regions of the spin Hall metal directly under the edges of the Cu overlayer where the inverse spin Hall charge current originates and ends.

To properly calculate the shunting correction factor in the Otani group experiments, one can perform a finite-element calculation for the electric potential $\varphi$ that maintains consistency with the continuity equation. If the spatial distribution of $J_{ISHE}$ is known, the governing equations to determine the backflow current density distribution $J_\varphi$ and the associated electric potential generated by the backflow take the form

$$\nabla \cdot (\vec{J}_\varphi + \vec{J}_{ISHE}) = 0 = \nabla \cdot (-\sigma \nabla \varphi + \vec{J}_{ISHE}) \tag{1}$$

or

$$\nabla^2 \varphi = \nabla \cdot \vec{J}_{ISHE} / \sigma . \tag{2}$$

(This is instead of $\nabla^2 \varphi = 0$ everywhere within the sample as assumed in ref. [13,14].) In other words, the electrical potential can be determined by solving the Laplace equation subject to the source term $\nabla \cdot \vec{J}_{ISHE} / \sigma$ and with the boundary conditions that the normal component of the total charge current density is zero on all the sample boundaries. For the relevant experimental geometry [13,14], $\vec{J}_{ISHE}$ has the form of a layer of current density (with thickness approximately $\lambda_{sf}$) within the spin Hall metal that has almost entirely a lateral component, and this component is to a good approximation constant as a function of the lateral coordinate except in the regions of the spin Hall metal directly under the edges of the Cu overlayer where the inverse spin Hall current originates and ends (the small shaded regions in Fig. 1(b)). Therefore the source term in the Laplace equation, $\nabla \cdot \vec{J}_{ISHE} / \sigma = (dJ^x_{ISHE}/dx)/\sigma$, is zero everywhere except in these shaded regions. The key point, for spin Hall materials with short values of $\lambda_{sf}$, is that volumes of these source regions are localized to within a distance of about $\lambda_{sf}$ from the interface with the low-resistivity Cu layer. This has the consequence, when $\lambda_{sf}$ is short, that a larger fraction of the backflow charge current travels via the low-resistivity Cu overlayer than in the Otani-group calculations, meaning that the voltage generated by the inverse spin Hall effect experiment (and hence the shunting correction factor $x_S$) is smaller than stated in references [13] and [14]. The actual backflow charge current density distribution that results for a material like Pt with a short spin diffusion length is plotted in Fig. 1(b). Note that this distribution is significantly different than the backflow charge current density pattern calculated by the procedure of references [13] and [14] (equivalent to Fig. 1(a)), particularly in the lower half of the spin Hall metal wire. To illustrate this point even more clearly, in Fig. 3 we plot the distribution of the lateral component of the charge current density just within a Ta layer near a Ta/Cu interface when current flows in response to a voltage applied to the ends of the Ta wire (Fig. 3(a)) compared to when a backflow current is generated by the inverse spin Hall effect assuming a 2 nm spin diffusion length (Fig. 3(b)). (Since Ta has a lower conductivity than Pt -- we assume $\sigma_{Ta}/\sigma_{Cu} = 0.005$ compared to $\sigma_{Pt}/\sigma_{Cu} = 0.13$ -- the importance of the shunting effects is enhanced relative to Pt.) The end result of the failure to obey the continuity equation in [13] and [14] is an underestimation of the spin Hall angle by an amount that depends on the spin diffusion length and resistivity of the spin Hall metal. Our finite-element calculations suggest that the spin Hall conductivities and spin Hall angles reported in ref. [14] are too low for Pt by a moderate amount (approximately a factor of 0.5) and for Ta by a much greater degree (approximately a factor of 0.1). These values are consistent with the corrections we determined separately for the direct spin Hall configuration (recall the result shown in Fig. 2). That is, the corrections we calculate for the direct spin Hall effect and the inverse spin Hall effect obey Onsager reciprocity, as they should.

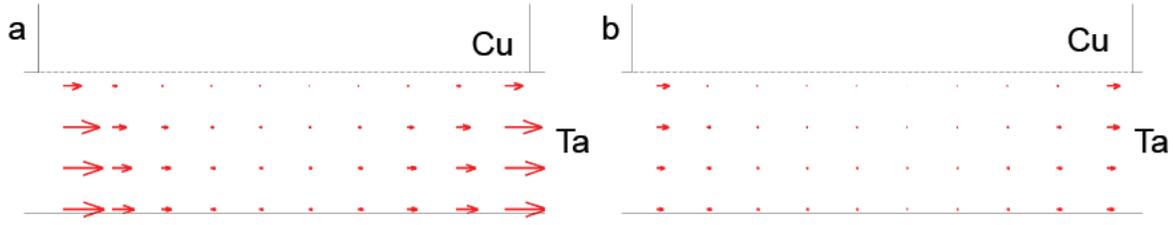

*Fig. 3*. *(a) Spatial distribution of just the lateral component of charge current density $J_x$ within just the Ta layer of a Cu/Ta device showing (a) the effects of current shunting by the Cu overlayer when a voltage is applied between the ends of the Ta wire (the $\nabla \cdot J_\phi = 0$ case) and (b) the backflow current generated by the inverse spin Hall effect when the spin diffusion length is 2 nm (the $\nabla \cdot J_\phi = -\nabla \cdot J_{ISHE}$ case). Note the difference in the current densities especially in the lower part of the Ta wire. We use $\sigma_{Ta}/\sigma_{Cu} = 0.005$. We assume that the Ta layer is 20 nm thick and the Cu overlayer is 100 nm long and 100 nm thick.*

The second experiment to report a quantitative measurement of the spin Hall angle $\theta_{SH}$ was by Ando *et al.* of the Saitoh group [17]. They measured the spin current generated by the spin Hall effect in a Pt/permalloy bilayer (with Pt conductivity $\sigma_{Pt} = 6.4 \times 10^6 \, (\Omega m)^{-1}$) by determining the magnitude of the spin transfer torque that it applies to the permalloy, via the effect of a dc spin transfer torque on the effective magnetic damping of the permalloy. Specifically, they measured the charge-current dependence of the effective magnetic damping in a magnetic-field-driven ferromagnetic resonance experiment, finding $\sigma_{SH} = 5.1 \times 10^5 [\hbar/(2e)](\Omega m)^{-1}$ and $\theta_{SH} \approx 0.08$. We agree that this is an accurate strategy for determining $\theta_{SH}$. However, we differ with ref. [17] regarding three steps in the analysis of the experiment: (i) Reference [17] used an approximation for the dependence of the effective magnetic damping on the spin current that can be written using our definitions of the experimental variables (and assuming that the equilibrium angle between the magnetic moment and the direction of the charge current is 90°) in the form $\Delta\alpha \approx \frac{\hbar J_S \gamma}{2eM_S t 2\pi f} = \frac{\hbar J_S}{2eM_S t \mu_0 [H_{ext}(H_{ext} + M_{eff})]^{1/2}}$. Here $\gamma$ is the gyromagnetic ratio, $M_S$ is the saturation magnetization of the magnetic layer, $t$ is the thickness of the magnetic layer, and $f$ is the FMR frequency. However we believe that it is more accurate for an in-plane-polarized ferromagnetic layer to write [18,19] $\Delta\alpha = \frac{\hbar J_S}{2eM_S t \mu_0 [H_{ext} + (M_{eff}/2)]}$. (ii) In their analysis, Ando *et al.* assumed that the component of the spin current that applies a spin torque to the permalloy (the component perpendicular to the permalloy magnetization) would penetrate into the permalloy a distance governed by the spin diffusion length in permalloy, $\lambda_{Py} \approx 3$ nm. However, we believe that this spin component will not penetrate into the permalloy any significant distance because it will undergo precession on approximately the scale of the lattice spacing [20], so that effectively one should set $\lambda_{Py} = 0$ within the equations in ref. [17]. (iii) Ando *et al.* assumed a value for the room-temperature spin diffusion length in Pt of $\lambda_{sf} = 7$ nm for Pt with a conductivity $\sigma_{Pt} = 6.4 \times 10^6$ ($\Omega$m)$^{-1}$. The appropriate value for the spin diffusion length is a topic to which we will return repeatedly in this review. As we will discuss in detail near the end of this manuscript, we have measured $\lambda_{sf} = 1.4 \pm 0.3$ nm in thin Pt layers with $\sigma_{Pt} = 5 \times 10^6$ ($\Omega$m)$^{-1}$ at room temperature. Nevertheless, the cumulative impact of these three differences on the quantitative determination of $\theta_{SH}$ by the method of Ando *et al.* is relatively modest, resulting in an overestimate of $\theta_{SH}$ of order of 35%. We have repeated a version of the Ando *et al.* FMR linewidth measurement as one part of ref. [18], finding originally by our analysis (before we measured $\lambda_{sf}$) a rough lower bound that $\sigma_{SH} > 2.4 \times 10^5 [\hbar/(2e)](\Omega m)^{-1}$ and $\theta_{SH} > 0.048 \pm 0.007$. Using $\lambda_{sf} = 1.4 \pm 0.3$ and $t_{Pt}$ = 6 nm, together with the formula $J_S(t_{Pt})/J_S(\infty) = 1 - \text{sech}(t_{Pt}/\lambda_{sf})$ for the dependence of the spin current magnitude on the Pt thickness [18], we can refine this analysis now to give a definite value for the ratio of the spin current density transmitted from a bulk Pt sample to permalloy divided by the charge current density in the Pt, $J_S(\infty)/J_e = 0.05 \pm 0.01$ by this technique. If the Pt/permalloy interface is transparent for spin transport, then the spin Hall angle $\theta_{SH}$ has this same value, $0.05 \pm 0.01$, and $\sigma_{SH} \approx 2.5 \times 10^5 [\hbar/(2e)](\Omega m)^{-1}$. If there is any substantial reflection at the interface, then $\theta_{SH}$ is necessarily greater than $J_S(\infty)/J_e$.

A third type of quantitative measurement of $\theta_{SH}$ was performed by Mosendz *et al.* at the Argonne National Laboratory [21,12]. They used the magnetic field from an oscillating current in a Au coplanar waveguide situated above (and electrically insulated from) a Pt/permalloy bilayer wire (with $\sigma_{Pt} = 2.4 \times 10^6 (\Omega m)^{-1}$) to drive magnetic resonance in the permalloy, and measured a dc voltage signal along the Pt/permalloy bilayer generated by a combination of two effects: (a) spin pumping together with the inverse spin Hall effect (ISHE) and (b) an anisotropic magnetoresistance (AMR) signal. They separated these two effects by arguing that the spin-pumping/ISHE signal should be strictly a symmetric function of applied field relative to the midpoint of the resonance peak, while the AMR signal should be strictly antisymmetric. In an initial analysis [21] they determined that $\sigma_{SH} = 1.6 \times 10^4 [\hbar/(2e)](\Omega m)^{-1}$ and $\theta_{SH}$ = 0.0067 for their Pt at room temperature, and their analysis was later refined [12] (by accounting for the effects of current shunting and the ellipticity of magnetic precession, although the arithmetic is somewhat unclear to us) to give $\sigma_{SH} = 3.1 \times 10^4 [\hbar/(2e)](\Omega m)^{-1}$ and $\theta_{SH} = 0.013 \pm 0.002$. Our primary criticism of this analysis is that Mosendz et al. assumed for the spin diffusion length a value $\lambda_{sf} = 10 \pm 2$ nm, referencing a measurement performed at 4.2 K by the Bass and Pratt group at Michigan State Univ. [22]. The Michigan State group quoted a value $\lambda_{sf}(4.2\text{ K}) = 14 \pm 6$ nm in Pt films with $\sigma_{Pt} = 24 \times 10^6 (\Omega m)^{-1}$. However, we wish to note that the Michigan State films had a very large conductivity compared to the material used by Mosendz et al., which had $\sigma_{Pt} = 2.4 \times 10^6 (\Omega m)^{-1}$ at room temperature [21]. It seems likely to us that the spin diffusion length will depend on both the temperature and the mean free path, so that $\lambda_{sf}$ may differ significantly between the Michigan State samples (low temperature and long mean free path) and the films used by Mosendz et al. (room temperature and shorter mean free path). If one makes the rough estimate that $\lambda_{sf} \propto \sigma_{Pt}$, as might be the case for spin relaxation dominated by the Elliot-Yafet mechanism [23], the value of $\lambda_{sf}$ appropriate for the samples of Mosendz et al. might be much smaller than 10 nm, presumably even smaller than our measurement $\lambda_{sf} = 1.4$ nm since the conductivity of the Pt films used by Mosendz et al. was half the value for our films. Just using our value for $\lambda_{sf}$, the analysis of Mosendz et al. would yield $\sigma_{SH} = 1.4 \times 10^5 [\hbar/(2e)](\Omega m)^{-1}$ and $\theta_{SH} \approx 0.06$, which is increased sufficiently relative to ref. [12] to be in better agreement with the measurements by other groups. (An even smaller value of $\lambda_{sf}$ would yield larger values for $\sigma_{SH}$ and $\theta_{SH}$.)

The analysis of Mosendz et al. has also been criticized by Azevedo et al. [24] and Harder et al. [25], both of whom argued that the AMR component of the signal need not be a strictly antisymmetric function of applied field relative to the center of the resonance because the oscillations of the magnetization and the induced current need not be exactly 90° out of phase at resonance as argued by Mosendz *et al.*, and therefore a more complicated analysis is required to separate the AMR and spin-pumping/ISHE contributions to the experimental signal. Both Azevedo *et al.* [24] and Harder *et al.* [25] show experimental evidence from FMR measurements on ferromagnet/nonmagnetic-metal bilayer devices that the phase between the current and the magnetization oscillations at resonance can be different than 90° and can change depending on the microwave frequency. However, their measurements were done using device geometries that were not identical to that of Mosendz *et al.* In contrast, Mosendz *et al.* reported that they conducted systematic control experiments as a function of changing their sample geometry and the microwave frequency [26], and also as a function inserting an insulating barrier between the Pt and permalloy layers in their sample [27], all of which support the argument that the phase between the current and magnetization oscillations in their device geometry is in fact 90° at resonance. Therefore they maintain that their analysis is correct as stated on this point.

We cannot provide a definite resolution for this argument here, but we can make some comments that may help to clarify the physics at issue. We will consider the generation of the AMR signal for a Pt/ferromagnet bilayer wire in the usual situation that the precession axis of the ferromagnet lies in the sample plane. We will assume that the Pt and the ferromagnet are in intimate electrical contact so that they carry a common current, with the oscillating local current densities in the ferromagnet and the Pt exactly in phase. We will also assume that precession in the ferromagnet is excited entirely by an in-plane oscillating magnetic field (with some component perpendicular to the magnet's precession axis), and that the spin transfer torque on the magnet arising from the spin Hall effect in the Pt layer can be neglected (we will say more about the case when this spin torque cannot be neglected when discussing the experiment of Liu *et al.* [18] below). If the device structure is such that the Oersted magnetic field generated by current flowing in the Pt/ferromagnet bilayer is the *only* source of oscillating magnetic field that is present to drive magnetic precession (as, *e.g.*, in the device structure of Liu *et al.* [18]), we believe that the physics should be simple. In that case the oscillating magnetic field is in phase with the current flowing in the Pt/ferromagnet bilayer wire. At the resonance condition both the magnetization response and the oscillating AMR resistance signal should be 90° out of phase from the magnetic field, so at resonance the AMR signal will also be exactly 90° out of the phase of the current. This gives an AMR signal that is an antisymmetric function of magnetic field about the resonance, in agreement with the discussion by Mosendz *et al.* However, the situation can be more complicated if there is an additional source of oscillating magnetic field, different from the current flowing in the Pt/ferromagnet wire. This is the case in the measurements of Mosendz *et al.* [12,21] (where the largest oscillating magnetic field is due to a gold coplanar waveguide above the Pt/ferromagnet wire) and in Azevedo *et al.* [24] (where the largest oscillating magnetic field is from a microwave resonant cavity). Now the AMR signal will depend on whether or not there is a relative phase between the total oscillating magnetic field acting to excite precession in the magnetic layer and the current in the Pt/ferromagnet wire that detects the AMR signal. Mosendz argue that there is no such phase difference in their device structure -- that the dominant source of oscillating current in their Pt/ferromagnet wire is capacitive coupling to the gold waveguide so that the currents in the two structures should be in phase. However, we suspect that this outcome may depend on details of the device structure, because capacitive coupling is not the only possible source of oscillating current in the Pt/ferromagnet wire. Inductive coupling between the waveguide and the Pt/ferromagnet wire will induce an electromotive force (emf) in the Pt/ferromagnetic wire that is phase shifted 90° from the current in the gold waveguide, and Faraday induction generated by the oscillating magnetization in the Pt/ferromagnet wire might also generate a phase-shifted emf. The phase and relative magnitude of the oscillating current that is generated within the Pt/ferromagnet wire by these inductive emfs will depend on details of the experimental geometry, in particular the value of the microwave-frequency complex impedance of the external circuit within which the Pt/ferromagnet wire is connected. This can be a complicated function of frequency, so in principle it is possible that the current in the Pt/ferromagnet wire might be in phase with the oscillating magnetic field at one frequency and substantially out of phase at a closely-neighboring frequency [24,25]. Like Azevedo *et al.* [24] and Harder *et al.* [25], we have observed these types of phase shifts with frequency, appearing as changes in the symmetry of the AMR signal with frequency, in devices with designs similar to (but not exactly the same as) Mosendz *et al.* The only conclusion we draw is that it is necessary to be careful about the microwave engineering in experiments like that of Mosendz *et al.* in order that the AMR resistance oscillations have a well-controlled phase relative to the current in the Pt/ferromagnet wire. Based on the analysis of Azevedo *et al.* [24] and the relative amplitudes of the field-symmetric and antisymmetric signals measured by Mosendz *et al.* [12], the differences in the final determination of the spin Hall angle $\theta_{SH}$ that result from disagreements about accounting for this phase appear to be less than a factor of two for Pt, so that this is a somewhat smaller discrepancy than the differences caused by uncertainties about the correct value for the spin diffusion length.

Azevedo *et al.* [24] carried through an independent analysis to determine the spin Hall angle using their analog of the Mosendz *et al.* spin-pumping/ISHE experiment. In the case of Azevedo *et al.*, magnetic resonance in a large-area Pt/permalloy sample ($\sigma_{Pt} = 2.42 \times 10^6$ $(\Omega\text{m})^{-1}$) was excited by an oscillating magnetic field in a resonant microwave cavity (not by a coplanar stripline as in the Mosendz *et al.* work), and the oscillating electromagnetic field of the cavity also generated an oscillating current within the Pt/permalloy bilayer. Azevedo *et al.* separated the AMR and spin-pumping/ISHE contributions to the resulting dc voltage by analyzing the signal as a function of rotating the sample angle relative to the cavity, instead of assuming a strictly antisymmetric-in-field lineshape for the AMR signal. They reported a value for the spin Hall angle $\gamma_H = 0.08 \pm 0.01$. However, it appears that their definition of the spin Hall angle may differ by a factor of two from ours ($\theta_{SH} = \gamma_H / 2$), which would mean that the analysis of Azevedo *et al.* would yield $\sigma_{SH} = 9.7 \times 10^4 [\hbar/(2e)](\Omega m)^{-1}$. Furthermore, the analysis of Azevedo *et al.* does not account properly for the shunting of the ISHE signal by the permalloy layer as explained by Nakayama *et al.* [11] and Mosendz *et al.* [12] (this causes an underestimate for $\gamma_H$ and $\sigma_{SH}$) and for their Pt films with $\sigma_{Pt} = 2.42 \times 10^6 \, (\Omega m)^{-1}$ Azevedo *et al.* used a value for the spin diffusion length of Pt equal to $3.7 \pm 0.2$ nm, larger than we believe is correct (which would cause a further underestimate of $\gamma_H$ and $\sigma_{SH}$). Azevedo *et al.* determined their value of the spin diffusion length by performing an analysis of the spin-pumping/ISHE signal magnitude as a function of changing Pt thickness, but we believe that this analysis contains an error because it neglected the shunting effect, and the correction factor required to account for shunting will vary depend on the thickness of the Pt layer.

Recently, the microwave-cavity based techniques have been improved by placing the sample in a position along the axis of a cylindrical cavity so that by the symmetries of the electromagnetic fields no oscillating current is induced within the Pt/ferromagnet bilayer and hence no AMR contribution is present [28-30]. Another signal that might arise in large-area samples from the anomalous Hall effect is also absent by virtue of the absence of any induced oscillatory current. Therefore the dc voltage outputs of these experiments are due entirely to the spin-pumping/ISHE effect, and the analysis of the signals is simplified. Ando *et al.* [30] have analyzed the signals from Pt/permalloy films ($\sigma_{Pt} = 2.0 \times 10^6$ $(\Omega m)^{-1}$) for this case using a procedure similar to that of the Mosendz *et al.* [12], taking into account the shunting effect. Assuming a value of the spin diffusion length in Pt of 10 nm and with a Pt layer thickness also 10 nm, Ando *et al.* calculated $\sigma_{SH} = 8 \times 10^4 [\hbar/(2e)](\Omega m)^{-1}$ and $\theta_{SH} = 0.04$. Like the result of Mosendz *et al.*, we believe that these are underestimates because of the large value of the spin diffusion length that was assumed. If one uses instead the value we have measured, $\lambda_{sf} \approx 1.4$ nm, the data of Ando *et al.* would suggest corrected values $\sigma_{SH} = 2.6 \times 10^5 [\hbar/(2e)](\Omega m)^{-1}$ and $\theta_{SH} \approx 0.13$.

Finally, a fourth technique to measure the strength of the spin Hall effect was introduced by our group in Liu *et al.* [18]. Here the device configuration is simply a Pt/permalloy bilayer ($\sigma_{Pt} = 5 \times 10^6$ $(\Omega m)^{-1}$) patterned into a microstrip line. One passes a microwave current through the stripline to generate precession in the permalloy and detects a dc voltage generated along the wire. This experimental design differs from that of Mozendz *et al.* in that the magnetic oscillations are generated entirely by the current flowing within the Pt/permalloy (there is no separate waveguide to generate a large oscillatory magnetic field). This has three important consequences. First, the Oersted field arising from the current in the wire is in phase with the current in the wire, so that there are no subtleties regarding the phase of the AMR contribution to the dc output signal generated by the Oersted field. Control experiments with Cu/permalloy samples demonstrated that the Oersted field in this device geometry generates a purely field-antisymmetric signal that can be used to calibrate the microwave current in the Pt (since the relationship between the Oersted field and the current in the Pt can be calculated exactly for the stripline geometry). Second, the ratio of the oscillatory magnetic field to the oscillatory current in the Pt/permalloy wire in the samples of Liu *et al.* is much smaller than in the experiment of Mosendz *et al.* (where the large oscillatory magnetic field was primarily generated by the separate Au waveguide), with the consequence that the ratio of the magnetic precession angle to the oscillatory current is small and hence the size of the spin-pumping/ISHE signal is negligible relative to the AMR signal in the samples of Liu *et al.* Third, again because the oscillating magnetic field is small relative to the current density in the Pt/permalloy wire, a spin transfer torque acting on the permalloy arising from the direct spin Hall effect within the Pt can be comparable to the Oersted field (which was not the case for Mosendz *et al.*), giving rise to spin-torque-driven ferromagnetic resonance (ST-FMR). This additional spin torque acts at a 90° angle relative to the Oersted torque and produces a contribution to the AMR signal that is symmetric about the resonance field. In the experiment of Liu *et al.*, therefore, the field-symmetric part of the resonant dc voltage output corresponds to an AMR signal resulting from the direct spin Hall torque (in contrast to a spin-pumping/ISHE output resulting from an Oersted-field torque in Mosendz *et al.*), while the field-antisymmetric part of the output in Liu *et al.* corresponds to the AMR signal arising from the Oersted field (this part is similar to the analysis of Mosendz *et al.*). As a result, in Liu *et al.* the spin current generated by the spin Hall effect in the Pt/permalloy samples can be determined accurately from the ratio of the field-symmetric and field-antisymmetric parts of the dc voltage output. This technique is similar in spirit to the experiment of Ando *et al.* [17], in that it measures a spin transfer torque arising from the direct spin Hall effect in order to determine the spin Hall angle, but by detecting the amplitude of magnetic precession excited by an oscillating spin transfer torque instead of a change in magnetic damping due to a dc spin transfer torque, we achieve better accuracy.

The experiment of Liu *et al*. also has the added virtue that it is in a sense self-calibrated. This is because the Oersted field that provides a net torque on the permalloy (Py) layer comes *only* from the part of the current that flows in the Pt, not the current in the Py. The Oersted field from the current flowing in the Py layer points in opposite directions above and below the midplane of the Py layer, while the Py is very thin and thus unable to support significant magnetization variations in the vertical direction. Therefore the effect of the Oersted field from the current flow in the Py averages to zero and does not produce any net torque that can cause the overall Py magnetization to precess. Because the *same* current density in the Pt gives rise to both the Oersted field acting on the Py and the spin Hall torque, the spin Hall torque can be determined accurately from the ratio of the symmetric and antisymmetric components of the resonance peak, without a need to separately estimate the current in the Pt by more approximate methods. The results of Liu *et al*. experiment were originally stated [18] as a lower bound on the spin Hall angle, $\theta_{SH} > 0.056 \pm 0.005$ corresponding to $\sigma_{SH} > (2.8 \pm 0.3) \times 10^5 [\hbar/(2e)](\Omega m)^{-1}$, because at that time we had not yet measured the room-temperature spin diffusion length in Pt, $\lambda_{sf}$. However, we have since measured a series of samples with different Pt thicknesses using the same technique, and this now allows us to determine more accurate values for both $\lambda_{sf}$ and $\theta_{SH}$, and not just a lower bound for $\theta_{SH}$.

The procedure we used to measure $\lambda_{sf}$ is as follows. A series of Pt($t_{Pt}$)/Py(4 nm) thin film samples ($\sigma_{Pt} = 5 \times 10^6$ $(\Omega m)^{-1}$) were grown and patterned into microstrip lines as in ref. [18], with Pt thicknesses $t_{Pt}$ varying from 1.5 nm to 10 nm, widths 1 - 20 μm, and lengths 5 - 200 μm. (The results for $\theta_{SH}$ were consistent, independent of width and length for a given layer structure.) A very thin, high-resistivity Ta layer was employed as a capping layer for all of the samples. The Pt films were sputtered at room temperature from a 99.95% pure Pt target using 2 mtorr Ar at a deposition rate of 0.1 nm/s. The base pressure for the sputtering system was less than $2 \times 10^{-8}$ torr. Using the same ST-FMR technique as in ref. [18], we measured the dc voltage output generated by applying an oscillating current through the microstrip line as a function of sweeping an applied magnetic field through the magnetic resonance condition. By fitting the measured FMR signal to a sum of symmetric and antisymmetric Lorentzians, we determined separately the magnitudes of the signals due to spin transfer from the spin Hall current and due to the Oersted field, respectively (Eq. (2) of ref. [18]). The antisymmetric signal was consistent with the value expected due to the Oersted field generated by the current in the microstrip line. By Eq. (3) of ref. [18], after correcting for a typographical error due to incomplete conversion from CGS to MKS units, we can calculate the ratio of the spin Hall current density transmitted from the Pt to the permalloy relative to the charge current density within the Pt as

$$\frac{J_{S,\text{rf}}}{J_{e,\text{rf}}} = \frac{S}{A}\frac{e\mu_0 M_S t_{Pt} d_{Py}}{\hbar}[1+(M_{\text{eff}}/H_{ext})]^{1/2}. \tag{3}$$

Here $S$ is the amplitude of the symmetric Lorentzian, $A$ is the amplitude of the antisymmetric Lorentzian, $M_S$ is the saturation magnetization of the permalloy layer, $d_{Py}$ the thickness of the permalloy layer, and $M_{\text{eff}}$ is the effective perpendicular demagnetizing field. We use $\mu_0 M_S = \mu_0 M_{\text{eff}} = 0.75$ Tesla ($M_S$ determined by magnetometer measurement of test samples is in good agreement with $M_{\text{eff}}$ determined by fitting to the FMR resonance field with the Kittel formula). The final result for the spin current $J_{S,\text{rf}}$ as a function of Pt thickness is plotted in Fig. 4(a). We observe that $J_{S,\text{rf}}$ is approximately constant for large Pt thicknesses and decreases abruptly with decreasing $t_{Pt}$ for small values.

This dependence of the spin current on the Pt thickness has the form that is expected within a drift-diffusion analysis [31] of the spin flow that incorporates the spin Hall effect. If one assumes that no spin current can penetrate out through the bottom surface of the Pt layer, there must arise a vertical gradient in the spin-dependent electron chemical potentials adjacent to the bottom surface (within a length scale given by the spin diffusion length $\lambda_{sf}$) that produces a counterflowing spin current to cancel the spin-Hall-generated spin current near the bottom surface. This reduces the magnitude of the spin-Hall spin current in films whose thickness is comparable to $\lambda_{sf}$. In the limit of a transparent interface between the Pt and the permalloy (so that the spin current density transmitted into the permalloy is equal to the spin current density incident on the interface from the Pt side, for spins perpendicular to the permalloy magnetic moment) within the drift-diffusion analysis the spin current density should have the thickness dependence [18]

$$\frac{J_S(t_{Pt})}{J_S(\infty)} = 1 - \text{sech}\left(\frac{t_{Pt}}{\lambda_{sf}}\right). \tag{4}$$

We find excellent agreement between the data in Fig. 4(a) and a one-parameter fit to this expression, with $\lambda_{sf}$ = 1.4 ± 0.3 nm. We have also analyzed the data in Fig. 4(a) using a generalization of Eq. (4) that allows for a reflection coefficient $R$ at the Pt/permalloy interface [32]. For non-zero values of $R$, the fits to the data in Fig. 4(a) correspond to values of $\lambda_{sf}$ even smaller than 1.4 nm, ranging down to 1.0 nm for $R = 0.9$.

As an independent check, we can also analyze the dependence of the Gilbert damping coefficient α on the Pt thickness (Fig. 4(b)). For each set of samples, we determined α from the linewidth of the FMR resonance $\Delta H$, according to $\alpha = \gamma \Delta H / (2\pi f)$. As seen in Fig. 4(b), we measure only a very weak dependence of $\alpha$ on $t_{Pt}$ in the range $t_{Pt} \geq 1.5$ nm. Within the theory of spin pumping [33], the presence of the Pt should increase the damping coefficient of the permalloy layer by an amount $\alpha(t_{Pt}) - \alpha(t_{Pt} = 0)$ above the intrinsic value, and this increase should depend on the ratio $t_{Pt}/\lambda_{sf}$ in the form:

$$\frac{\alpha(t_{Pt}) - \alpha(t_{Pt} = 0)}{\alpha(t_{Pt} = \infty) - \alpha(t_{Pt} = 0)} = \frac{1 + [\sqrt{\varepsilon}]^{-1}}{1 + [\sqrt{\varepsilon}\tanh(t_{Pt}/\lambda_{sf})]^{-1}} . \quad (5).$$

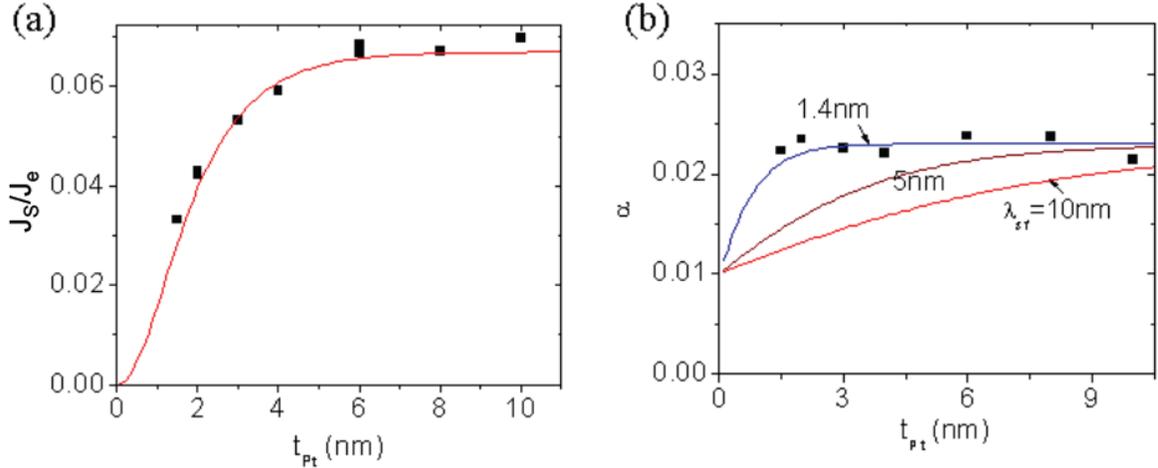

*Fig. 4. (a) Squares: Results from ST-FMR measurements for the spin current density transmitted from Pt to the permalloy layer, divided by the charge current density in the Pt, as a function of the Pt thickness. Red line: Fit to Eq. (4) that determines the value $\lambda_{sf} = 1.4 \pm 0.3$ nm. (b) Squares: The dependence of the damping coefficient α on $t_{Pt}$ as measured by the ST-FMR experiment. Curves: predictions of Eq. (5) for different values of $\lambda_{sf}$.*

This equation follows from Eq. (21) of ref. [33]. The intrinsic damping coefficient in our permalloy films with no coupling to a Pt layer has the value $\alpha(t_{Pt} = 0) \approx 0.01$, while the quantity $\varepsilon = \tau_{el}/\tau_{sf}$, the spin-flip probability at each scattering, is for Pt believed to be > 0.1 [33]. The red, brown and blue curves in Fig. 4(b) show the form of Eq. (5) for ε = 0.1 and $\lambda_{sf}$ = 10 nm, 5 nm, and 1.4 nm, respectively. The curves are only weakly dependent on the value assumed for ε; our conclusions hold for any ε in the range 0.1-1. The result of the comparing the data in Fig. 4(b) to Eq. (5) is that the weak dependence of Δα on $t_{Pt}$ implies a value of $\lambda_{sf}$ < 2 nm, fully consistent with the value we determined above from the dependence of the spin current density on $t_{Pt}$. A related measurement of magnetic damping as a function of the thickness of a sputtered Pt film was performed previously by Mizukami *et al*. for a Cu/permalloy/Cu/Pt/Cu multilayer structure (Fig. 4 of ref. [34]), from which one can draw a similar conclusion, that $\lambda_{sf} \approx 1$ nm at room temperature for Pt.

Returning to the results of the ST-FMR measurement, the data in Fig. 4(a) provide the means to extrapolate to large Pt thicknesses our measurements of the quantity $J_S/J_e$, the spin current density penetrating from Pt into the permalloy divided by the charge current density within the Pt. The best fit using the full set of samples is $J_S(\infty)/J_e = 0.068 \pm 0.005$. If the Pt/permalloy interface is transparent for spin transport, we can therefore conclude also that $\theta_{SH} = 0.068 \pm 0.005$ and $\sigma_{SH} = (3.4 \pm 0.3) \times 10^5 [\hbar/(2e)](\Omega m)^{-1}$. If there is any lack of transparency at the interface, these values represent lower bounds on the true values of $\theta_{SH}$ and $\sigma_{SH}$ within the Pt sample.

In summary, based on the arguments in this paper we can understand several reasons why different techniques for measuring $\sigma_{SH}$ and $\theta_{SH}$ at room temperature in Pt have found conflicting results. We have explained that the smallest reported values, determined by measuring either the direct spin Hall effect or the inverse spin Hall effect in lateral devices [7,14] are incorrect due to shunting of the high-resistivity Pt by a low-resistivity top Cu contact. In regard to other techniques, the published values of $\sigma_{SH}$ and $\theta_{SH}$ determined by exciting magnetic resonance in a Pt/permalloy bilayer and detecting a dc voltage generated by spin pumping and the inverse spin Hall effect [12,24,30] have generally been smaller than the values determined by experiments that measure the spin transfer torque generated by the direct spin Hall effect [17,18]. We suggest that the primary reason for this discrepancy is that the different analyses assumed different values for the spin diffusion length in Pt, $\lambda_{sf}$, in Pt/ferromagnet bilayer samples. The spin-pumping/ISHE experiments generally used values considerably larger than the value we have determined for our Pt layers at room temperature, $\lambda_{sf}$ = 1.4 ± 0.3 nm for our films with $\sigma_{Pt} = 5 \times 10^6 \, (\Omega m)^{-1}$. If our value for $\lambda_{sf}$ is applied to all of the experiments, the result is that the spin-pumping/ISHE experiments and the direct spin Hall effect/spin torque experiments give much better consistency than the original analyses. For the two spin-pumping/ISHE experiments for which we have enough information calculate a corrected value using our determination of $\lambda_{sf}$, we find $\sigma_{SH} = 1.4 \times 10^5 [\hbar/(2e)](\Omega m)^{-1}$ [12] and $\sigma_{SH} = 2.6 \times 10^5 [\hbar/(2e)](\Omega m)^{-1}$ [30], compared to the values from our direct spin Hall/spin torque experiments $\sigma_{SH} = 2.5 \times 10^5 [\hbar/(2e)](\Omega m)^{-1}$ and $\sigma_{SH} = (3.4 \pm 0.3) \times 10^5 [\hbar/(2e)](\Omega m)^{-1}$ [18]. Taking into account the differing conductivities of the Pt films, all of these values for the spin Hall conductivity correspond to large spin Hall angles, with $\theta_{SH}$ > 0.05. The remaining differences between the measurements might represent real differences in materials preparation (*e.g.,* differences in $\lambda_{sf}$, the dominant scattering processes that affect $\sigma_{SH}$, or the extent of intermixing at the Pt/permalloy interface) or additional unresolved issues in the analyses (*e.g.*, the controversy [24,25] over the phase between the magnetization oscillations and the induced current in the measurements of ref. [12]). Values of $\theta_{SH}$ greater than 0.05 are sufficiently large that the spin current arising from the spin Hall effect can be used to apply a strong spin torque that can efficiently rotate and even switch the magnetic moment in a ferromagnetic layer adjacent to the Pt [35].

Finally, we note that additional measurements would be useful to confirm or disprove our arguments that (i) $\lambda_{sf} \approx$ 1.4 nm in Pt at room temperature within Pt/ferromagnet bilayers, and (ii) that it is necessary to use a spin diffusion length on this scale to achieve consistency between the different techniques for measuring the spin Hall conductivity. In particular, spin-pumping/ISHE experiments using a sequence of Pt thicknesses comparable to $\lambda_{sf}$ should be able to provide an additional independent measurement of both $\lambda_{sf}$ and $\sigma_{SH}$ without making any assumptions about unmeasured variables. We suspect that it might be possible to reanalyze the existing data of Azevedo *et al*. [24], taking into account the shunting effect that was neglected originally, in order to achieve this goal. Once consensus is reached about how to interpret the various different types of spin Hall experiments, we anticipate that the field will be able to move forward to understand more fundamental questions such as whether the origin of the spin Hall effect in particular materials is due to intrinsic or extrinsic mechanisms, how the $\sigma_{SH}$, $\theta_{SH}$, and $\lambda_{sf}$ vary as a function of doping and temperature, and how the strength of the spin Hall effect might be optimized to enable efficient spin-torque manipulation of magnetic devices.

Notes added:

(1) H. Nakayama *et al*. have a preprint in which they analyze a spin-pumping/ISHE experiment as a function of Pt thickness [36].

(2) Prof. Otani has communicated to us that he believes strongly that the spin diffusion length in the Pt elements within his group's nonlocal spin valve devices is not as short as we have measured in our Pt/ferromagnet bilayers at room temperature. He reports that his group has performed weak antilocalization measurements (to be published) on pure Pt wires from which they have determined a spin diffusion length of approximately 12 nm at 3 K, consistent with their results from nonlocal spin absorption measurements in ref. [14].

Acknowledgments: We thank many colleagues for communications on this topic, particularly K. Ando, A. Azevedo, G. E. W. Bauer, A. Fert, A. Hoffmann, C.-M. Hu, and Y. Otani. This thanks is not meant to imply that they necessarily endorse all of our conclusions. We acknowledge support from ARO, ONR, NSF/MRSEC (DMR-1120296) through the Cornell Center for Materials Research (CCMR), and NSF/NSEC through the Cornell Center for Nanoscale Systems. We also acknowledge NSF support through use of the Cornell Nanofabrication Facility/NNIN and the CCMR facilities.

Also note that Eq. (4) in our paper ref. [18] contains a typographical error resulting from an incomplete conversion from CGS to MKS units: $2\pi M_{eff}$ should be $M_{eff}/2$ in Eq. (4) of ref. [18], but the final quantitative results in ref. [18] were calculated correctly.)